\documentstyle[twocolumn,aps]{revtex}

\begin{document}

\draft

{\bf Buhot Reply :} 
In the letter~\cite{Buhot} I proposed a possible scenario 
for the phase-separation instability of binary mixtures of 
hard-core particles in the limit of high asymmetry between 
large and small particles relating this transition to a 
bond-percolation transition. 
In his Comment~\cite{Louis} A. A. Louis claims that, at least 
in the case of hard-spheres, the phase-separation transition 
is {\it unrelated} to bond-percolation. In fact, the mapping 
to bond-percolation relates to an instability of the homogeneous 
phase and, in the case of hard-spheres (or hard-disks) it gives 
only an upper bound for the packing fraction at the phase 
separation transition. 

First of all, let me recall the mapping argument to the 
bond-percolation transition proposed in~\cite{Buhot}. 
Due to the large asymmetry, 
the radial distribution function $g_{ll}(r)$ of the large 
particles possesses a sharp and high peak of width $\sigma_s$ 
(size of small particles) at the contact value $\sigma_l$ 
between two large particles. 
This peak may be interpreted as {\it bonds} between large 
particles and the number of bonds $n_b$ is then defined as:

\begin{equation}
n_b = \rho_l \int_{\sigma_l \leq r \leq \sigma_l + \sigma_s}
g_{ll} (r) d{\bf r}
\end{equation}

\noindent where $\rho_l$ is the number density of the large 
particles. As the number $n_b$ increases with the total 
packing fraction, larger and larger aggregates of large 
particles appear in the {\it homogeneous} fluid phase. 
For a sufficiently large number of bonds $n_c$ or 
equivalently a sufficiently large packing fraction, 
a macroscopic aggregate appears in the system.  
This macroscopic aggregate breaks the translational 
invariance of the system which is no more homogeneous.  
Thus, the homogeneous fluid phase is instable. 
In the letter~\cite{Buhot} I approximate the number 
$n_c$ by the number of bonds $z p_c$ at the bond-percolation 
transition of the corresponding crystal lattice of large 
particles where $z$ is the coordination number and $p_c$ 
the bond-percolation threshold. This approximation  
neglects two effects : the slight modification due to the 
lack of lattice and, more important, the possible 
correlations between bonds.   
Taking into account those possible correlations, 
the appearance of the solid phase (or macroscopic aggregate)
will be shift to a lower number of bonds ($n_c < z p_c$).
The packing fraction $\eta_c$ corresponding to $n_b = z p_c$ 
is thus only an upper bond of the packing fraction $\eta_t$ 
at the phase-separation transition between the fluid phase 
and the fluid-solid phase.     

In the case where the phase-separation transition is 
strongly first order due to a large surface tension 
between the fluid and the solid, we may expect large
correlations between the bonds and then the bond-percolation 
transition is not directly related to the phase-separation 
transition. However, $\eta_c$ is an upper bound (and thus 
only a qualitative estimation) for $\eta_t$. 
This is exactly what is observed on Fig.1 of the 
Comment~\cite{Louis} since, for hard-spheres, we expect a 
large surface tension.

It is also interesting to notice that for hard-disk mixtures 
the prediction that instability of the homogeneous fluid phase
occurs for sufficiently high asymmetry remains valid.  
Consequently, a phase-separation transition is  
predicted as already said in~\cite{Buhot}.
However, no simulations are yet able to confirm this result. 

Concerning the case of parallel hard-cubes, as already said
in the Comment~\cite{Louis}, 
it is known that the freezing of the one component fluid is 
a second order transition~\cite{Jagla}. This is principally 
due to the lack of rotational symmetry since the cubes are 
parallel~\cite{Jagla,Cuesta} (the first order nature of the  
freezing transition is restored if we allow the cubes to rotate). 
Thus, in the case of binary mixtures of parallel hard-cubes 
(or squares), we may expect that the surface tension between
the fluid and the solid is low. In that case, we may expect
that the packing fraction $\eta_c$ is a quantitative 
approximation for $\eta_t$. Numerical simulations for few 
state points seem to confirm this result but a complete 
numerical calculation of the phase diagram is still lacking.   

In conclusion, the phase-separation transition is not {\it directly} 
related to bond-percolation transition. However, the mapping
proposed in~\cite{Buhot} may be useful to predict the existence
of phase separation in binary mixtures of hard-core particles
since it gives an upper bound of the packing fraction at the
phase separation transition. 
It is for example the case for the hard-disks mixture.

\vskip 0.5cm
\noindent A. Buhot

Theoretical Physics

University of Oxford

1 Keble Road

Oxford OX1 3NP

UNITED KINGDOM

\bibliographystyle{unsrt}

\begin{thebibliography}{99}

\bibitem{Buhot}
A. Buhot {\it Phys. Rev. Lett.} {\bf 82}, 960 (1999).

\bibitem{Louis}
A. A. Louis preceding Comment.

\bibitem{Jagla}
E. A. Jagla {\it Phys. Rev. E} {\bf 58}, 4701 (1998).

\bibitem{Cuesta}
Y. Martinez-Rat\'on and J. A. Cuesta {\it J. Chem. Phys.} 
{\bf 111}, 317 (1999).

\end{thebibliography}

\end{document}